# CAUGHT IN THE ACT: THE IDENTIFICATION OF THE GALAXIES RESPONSIBLE FOR THE FAINT BLUE EXCESS

SIMON P. DRIVER AND ROGIER A. WINDHORST

*Arizona State University,*
*Department of Physics and Astronomy,*
*Box 871504, Tempe, AZ85287-1504, USA*

AND

RICHARD E. GRIFFITHS

*Bloomberg Center for Physics and Astronomy, JHU*

**Abstract.** We summarise recent Hubble Space Telescope results on the morphology of faint field galaxies. Our two principle results are: (1) the galaxies responsible for the faint blue excess have late-type/irregular morphology and (2) the number counts of normal galaxies, ellipticals and early-type spirals, are well fit by standard no-evolution models implying that the giant population was in place and mature by a redshift of $\geq 0.7$.

## 1. Introduction

One of the great leaps forward in the study of faint field galaxies, and made possible only by the Wide Field and Planetary Camera 2 (WFPC2) on board the Hubble Space Telescope (HST), is the ability to discern detailed morphological information to faint magnitudes ($I \leq 24.24$). The Medium Deep Survey (a key HST project, c.f. Griffiths *et al.* 1994) has exploited this new area resulting in a number of recent morphology related publications (Casertano *et al.* 1995; Driver, Windhorst & Griffiths 1995; Driver *et al.* 1995; Glazebrook *et al.* 1995). Here we briefly summarise and coalate some of the key results from these papers, and interpret them by comparison to the predictions of some generic faint galaxy models. We also discuss the redshift dsitribution for Late-type/Irregulars which can now be directly derived based on the mounting observational evidence that the "giant" galaxies are not strongly evolving over the redshift range $0.0 < z < 1.0$ (c.f. Driver *et al.* 1995; Mutz *et al.* 1994; Lilly; Dickinson these proceedings).

## 2. Summary of the MDS morphology papers

CASERTANO *et al.* 1995 - Casertano analysed the entire WF/PC database (13,500 galaxies), using an automated technique to classify the sample into bulge dominated (E/S0), disk Dominated ($\sim$ Sabcd) or other systems (M/Irr/?). The principle results are that at faint magnitudes ($20 < I_{785LP} < 21$) the number counts are dominated by very small disk systems with a mean scale-length of $\sim 0.3$ arcseconds.

DRIVER, WINDHORST & GRIFFITHS 1995 - A complete magnitude limited sample ($I_{814} < 22$) of 144 field galaxies drawn from six MDS fields. Classifications are by eye into ellitpicals (E/S0), early-type spirals (Sabc) and late-type spirals/Irregulars (Sd/Irr). At $I_{814} \sim 22.0$ mag, these three populations are observed in almost equal proportions. While the E/S0's and Sabc's are well fit by "conventional" models, the Sd/Irr's have about 1 dex higher surface densisty than expected.

GLAZEBROOK *et al.* 1995 - A similar analysis to that above was made independently based on a sample of 301 galaxies drawn from 10 MDS fields over a comparable magnitude range. The findings are similar demonstrating a robustness to the observational result.

DRIVER *et al.* 1995 - The most recent survey based on a single *ultradeep* HST field (now totalling 67 orbits, c.f. Windhorst - this conference; and Windhorst & Keel 1995). The results confirmed the earlier work and find that the galaxy counts beyond $I814 = 22$ continue to become increasingly dominated by Sd/Irrs. (see colour plate by Windhorst, these proceedings).

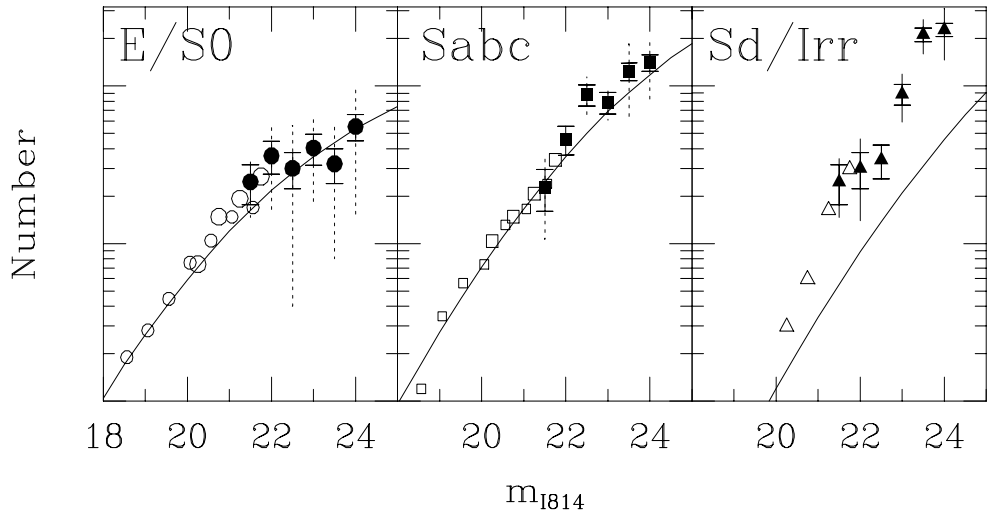

*Figure 1.* A compendium of morphological galaxy counts compared to the "standard" no-evolution models. Note that the models are renormalized at $b_J \sim 18$ mag, which is somewhat naive as the justification — local evol$^n$, local hole, SB sel$^c$ effects and/or photometric effects — are liable to be both type and luminosity dependent.

## 3. Popular Generic Faint Galaxy Models

Recent popular models to explain the Sd/Irr counts are:

DWARF MODELS, e.g. Driver *et al.* (1994) - These exploit the uncertainty in the space density of Sd/Irrs and increase the normalization and slope of the Sd/Irr LF until an optimal fit is found. Such models give good fits to the counts in all bands and can explain the trend to bluer colours at fainter magnitudes. The models typically fail to match the z-distribution.

EVOLVING FLAT MODELS, e.g. Broadhurst, Ellis & Shanks (1988) - Assumes the LF for Sd/Irrs is an extrapolation of the luminous galaxies, then luminosity evolution is invoked to match the counts. The models match the z-distributions but the luminosity evolution required to match the counts is extreme, $\sim 2.5$ mags in every Sd/Irr galaxy by z=0.5.

MERGER MODELS, e.g. Broadhurst, Ellis & Glazebrook (1992) - As above except merging is invoked. The merger rate required is extreme and not seen in $w(\theta)$. The morphological data also fail to show the density of close companions required. That the E/S0 and Sabc galaxies fit closely to the no-evolution models argues against merging, since if these were the merger products, they should be significantly rarer in the past.

AN EVOLVING DWARF MODEL, e.g. Phillipps & Driver (1995) - Essentially taking the best of the first two models its clear to see that a good half-way solution should be found. The data for Sd/Irrs show $\sim 60$ % inert, $\sim 20$ % tidally disrupted and $\sim 20$ % knotted cores (spontaneous starbursts ?), supporting all the above models to some extent. This model, although contrived, can fit all the counts and the z-distributions.

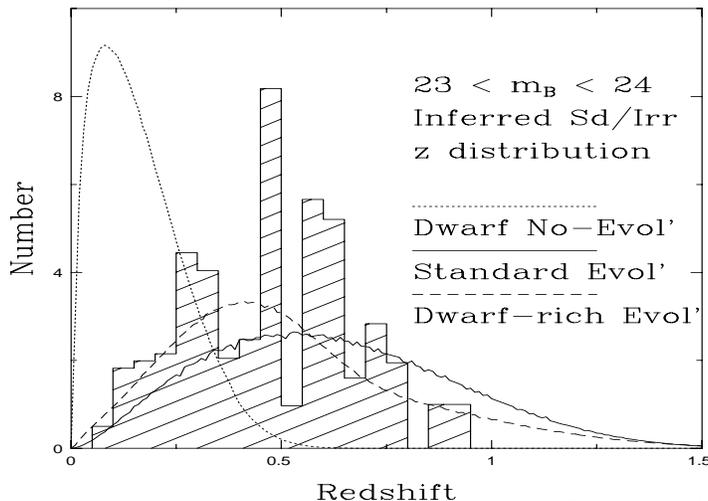

*Figure 2.*   A comparison of the above models to the inferred Sd/Irr redshift distribution. This distribution is derived by subtracting off the predicted distribution for E/S0 and Sabc, assuming no evolution as implied by their counts, from the total redshift distribution of Glazebrook *et al.* (1995).

## 4. Discussions, Conclusions and Speculations

Perhaps the most far reaching gain, is that now the culprits responsible for the faint blue excess have been isolated, we can return to the original aim of number counts, namely constraining cosmological models. With the mounting evidence that the E/S0 and Sabc galaxies are largely in place by $z \sim 1$, this aim becomes even more attainable. Already the counts presented here rule out any posibility of a large comsological constant (c.f. Driver *et al* 1995) and thereby block the possiblity of using $\Lambda$ as a way out of the age of the universe problem. Nevertheless such analaysis is still fraught with perils and most surprisingly, it is our knowledge of the local environment which provides the largest remaining obstacles.

**Questions**
WHITE: You showed us that your E/S0 counts are consistent with *no* evolution in an Einstein-de Sitter universe. Are they also consistent with no evolution in an open universe or with passive evolution in either ?
DRIVER: We've explored all of these alternatives and conclude that enough margin of error exists that we cannot rule for or against any of the models you mentioned, yet. We can however rule out a $\Lambda$-dominated universe.
WORTHEY: Do you have a histogram of morphological type versus colour.
DRIVER: Yes, but we find little or no correlation between type and colour, other than that few late-types are seen at red colours. This may be due to the poor range of the colour baseline ($V_{606}$ and $I_{814}$).